\documentstyle[12pt]{article}
\input epsf.sty
\newcommand{\be}{\begin{equation}}
\newcommand{\ee}{\end{equation}}
\newcommand{\bea}{\begin{eqnarray}}
\newcommand{\eea}{\end{eqnarray}}
\newcommand{\al}{\alpha}

\newcommand{\eps}{\epsilon}

\newcommand{\th}{\theta}

\newcommand{\lm}{\lambda}

\newcommand{\nn}{\nonumber}

\begin{document}
\title{Strings in Kerr-Newmann Black Holes.}
\author{A. Kuiroukidis\thanks{E-mail: kuirouki@astro.auth.gr} 
and D. B. Papadopoulos\thanks{E-mail: papadop@astro.auth.gr} \\
Department of Physics, \\
\small Section of Astrophysics, Astronomy and Mechanics, \\
\small Aristotle University of Thessaloniki, \\
\small 54006 Thessaloniki, GREECE}

\maketitle
\begin{abstract}
We study the evolution of strings in the 
equatorial plane of a Kerr-Newmann black hole. 
Writing the equations of motion and the constraints resulting 
from Hamilton's principle, three classes of 
exact solutions 
are presented, for a closed string, encircling the black-hole. 
They all depend on two arbitrary, integration functions and two constants. 
A process for extracting energy is examined for the case 
of one the three families of solutions. This is the analog of the 
Penrose process for the case of a particle.

\end{abstract}

\section*{I. Introduction}\

In modern physics, string theory plays a prominent role 
in the effort to achieve a consistent quantization of 
gravity and to provide a unified description of all the fundamental 
interactions [1-2]. 
String theory has overcome several longstanding problems in high
energy 
physics in a beautiful and elegant way and has given the most convincing 
perspective in the unification effort. So the formulation of the theory in 
a background that contains gravity has been recognized from the early days as 
neccesary for extracting valuable information for its content.  
This is the strongest motivation for the study of strings in curved 
spacetimes [2-3]. 

In the context of Black-Hole physics, strings have been used as candidates
for the resolution of the main paradoxes and unsolved problems 
associated with it, and effective string actions in background fields 
have been investigated [4-5].

The string equations  in the Kerr-Newmann spacetime are highly non-linear
and difficult to solve exactly. Various attempts have been made to reduce 
them using different kinds of ansatze [6-7], 
by considering the Nambu-Goto action or by studying generic 
null-string configurations [8-11]. 
The problem is simplified if strings are bound to move on the equatorial plane 
of the Kerr-Newmann black hole. One can consider then the main problems of 
black hole physics in the light of string theory. The aim of this paper is 
to provide a method of obtaining 
three families of solutions and to examine the process of energy 
extraction for one of these families. 

This paper is organized as follows:

In section II, the general features of string theory in the Kerr-Newmann
black-holes are given.

In section III, the equations of motion for strings in the equatorial plane 
of the Kerr-Newmann black-hole are given.

In section IV, three classes of solutions are presented and analyzed.

In section V, the process for extracting energy from the black-hole
via strings is examined for a special case.

\section*{II. Strings in Black-Hole Spacetimes}\
The action for a bosonic string in a curved-spacetime background is given 
(for a D-dimensional spacetime) by [1,2] 
\be 
S \: = \: -\frac{T}{2} \ \int d \tau d \sigma \sqrt{-h}h^{\alpha \beta }(\tau ,\sigma )
G_{MN}(X)\partial _{\alpha}X^{M} \partial _{\beta}X^{N} 
\ee 
where $ M,N = 0,1,...,(D-1) $ are spacetime indices, $ \alpha ,\beta = 0,1 $
are worldsheet indices and $ T = (2\pi \alpha ^{\prime })^{-1} $ is the 
string tension.

Variation of the action with respect to the "fields" which are the string 
coordinates $ X^{M}(\tau ,\sigma )$, gives the equations of motion and the
constraints [2-3]
\bea 
\ddot{X}^{M}-(X^{M})^{\prime \prime}+\Gamma ^{M}_{AB}[\dot{X}^{A}\dot{X}^{B}
-X^{\prime A}X^{\prime B}] = 0 \\
G_{AB}(X)[\dot{X}^{A}\dot{X}^{B}+X^{\prime A}X^{\prime B}] = 0 \\ 
G_{AB}(X)\dot{X}^{A}X^{\prime B} = 0 
\eea 
where the dot stands for $ \partial _{\tau } $ and the prime stands for
$ \partial _{\sigma } $. 
In the conformal gauge choice, $ h_{\alpha \beta }(\tau ,\sigma ) =
exp[\Sigma (\tau ,\sigma )]\eta _{\alpha \beta }$
where $ \Sigma (\tau ,\sigma ) $ is an arbitrary function and 
$\eta _{\alpha \beta }=diag(-1,+1)$
we have the Lagrangian density 
\be 
{\cal L} =G_{MN}(X)[-\dot{X}^{M}\dot{X}^{N}+X^{'M}X^{'N}] 
\ee 
from which the equations of motion follow via the Euler equations
\be 
\partial _{\alpha }\left( \frac{\partial {\cal L}}
{\partial (\partial _{\alpha }X^{M})}\right)-
\frac{\partial {\cal L}}{\partial X^{M}} = 0 
\ee 
This constitutes a two-dimensional field theory. 
The equations of motion and the constraints for null strings are 
given by [4]
\bea 
\ddot{X}^{M}+\Gamma ^{M}_{AB}\dot{X}^{A}\dot{X}^{B}= 0 \\
G_{AB}(X)\dot{X}^{A}\dot{X}^{B}= 0 \\ 
G_{AB}(X)\dot{X}^{A}X^{\prime B} = 0 
\eea 
For the case of the four-dimensional Kerr-Newmann black hole 
we have the string coordinates $(X^{0},X^{1},X^{2},X^{3})$, 
corresponding to the coordinates $(t,r,\theta ,\phi )$, of 
the Boyer-Lindquist coordinate system.

We make the definitions
\bea 
E\equiv r^{2}+\alpha ^{2}+Q^{2}-2Mr \\
\Delta \equiv r^{2}+\alpha ^{2}cos^{2}\theta \\
\delta \equiv r^{2}+\alpha ^{2}\\ 
\epsilon \equiv r^{2}-\alpha ^{2}cos^{2}\theta  
\eea 
where Q is the charge of the black hole, M is its mass and
$\alpha \equiv S/M$ is the angular momentum per unit mass.

The metric is given by
\bea 
ds^{2}&=&-\frac{E}{\Delta }[dt-\alpha sin^{2}\theta d\phi ]^{2}
+\frac{sin^{2}\theta }{\Delta }
[\delta d\phi -\alpha dt]^{2}+\nn \\
&+&\frac{\Delta }{E}dr^{2}+\Delta d\theta ^{2} 
\eea   

The static limit is given by [12,13]
\be 
S_{\pm }:r=M\pm \sqrt{M^{2}-Q^{2}-\alpha ^{2}cos^{2}\theta } 
\ee 
The horizon is given by
\be 
\Sigma _{\pm }:r=M\pm \sqrt{M^{2}-Q^{2}-\alpha ^{2}} 
\ee 
where we assume that $\; \; M^{2}\geq Q^{2}+\alpha ^{2} \; \; $.

The open string boundary conditions demand
\be 
(X^{M})^{'}(\tau ,\sigma =0)=(X^{M})^{'}(\tau ,\sigma =\pi )=0 
\ee 
while for closed strings we must have
\be 
X^{M}(\tau ,\sigma =0)=X^{M}(\tau ,\sigma =2\pi ) 
\ee 
\section*{III. The Equations of Motion}\
The worldsheet light-cone variables are defined by 
$\; \; \chi ^{\pm }=(\tau \pm \sigma )\; \; $. 
The Jacobian of the transformation is given by
\be 
J\equiv \frac{\partial (\chi ^{+},\chi ^{-})}
{\partial (\tau ,\sigma )}=-2,\; \; \; \; \; \; \; \;  
d\chi ^{+}d\chi ^{-}=-2d\tau d\sigma  
\ee 
We introduce the notation and the ansatz 
\bea 
X^{0}(\tau ,\sigma )=t(\tau ,\sigma ),
\; \; \; 
X^{r}(\tau ,\sigma )=r(\tau ,\sigma )\\ 
X^{\theta }(\tau ,\sigma )=(\pi /2),
\; \; \; 
X^{\phi }(\tau ,\sigma )=\phi (\tau ,\sigma ) 
\eea 
Defining 
\be 
R(r)\equiv (1-\frac{2M}{r}+\frac{Q^{2}}{r^{2}}) 
\ee 
\\
the Lagrangian density is given by
\bea
\bar{\cal L}&=&4R(\partial _{+}t)(\partial _{-}t)-4\alpha (R-1)
[(\partial _{-}t)(\partial _{+}\phi )+
(\partial _{+}t)(\partial _{-}\phi )]\nn \\&-&
4[r^{2}+2\alpha ^{2}-\alpha ^{2}R]
(\partial _{+}\phi )(\partial _{-}\phi )-
\frac{4r^{2}}{\alpha ^{2}+Rr^{2}}
(\partial _{+}r)(\partial _{-}r) 
\eea
The action is given by
\be 
S=\int {\cal L}d\tau d\sigma =\int -(\bar{\cal L}/2)d\chi _{+}d\chi _{-} 
\ee 
The equations of motion
\be 
\frac{\delta \bar{\cal L}}{\delta t}=
\frac{\delta \bar{\cal L}}{\delta r}=
\frac{\delta \bar{\cal L}}{\delta \phi }=0 
\ee 
read 
\be 
\partial _{+}[R(\partial _{-}t)-\alpha (R-1)(\partial _{-}\phi )]+
\partial _{-}[R(\partial _{+}t)-\alpha (R-1)(\partial _{+}\phi )]=0 
\ee 
\bea
\partial _{+}[(r^{2}+2\alpha ^{2}-\alpha ^{2}R)(\partial _{-}\phi )+
\alpha (R-1)(\partial _{-}t)]&+&\nn \\ 
+\partial _{-}[(r^{2}+2\alpha ^{2}-\alpha ^{2}R)(\partial _{+}\phi )+
\alpha (R-1)(\partial _{+}t)]&=&0 
\eea
\bea 
\partial _{+}[\frac{r^{2}}{\alpha ^{2}+Rr^{2}}(\partial _{-}r)]+
\partial _{-}[\frac{r^{2}}{\alpha ^{2}+Rr^{2}}(\partial _{+}r)]&=&\nn \\ 
\frac{\partial }{\partial r}[\frac{r^{2}}{\alpha ^{2}+r^{2}R}]
(\partial _{+}r)(\partial _{-}r)+
\frac{\partial }{\partial r}[r^{2}+2\alpha ^{2}-\alpha ^{2}R]
(\partial _{+}\phi )(\partial _{-}\phi )+\nn \\ 
+\alpha (\frac{\partial R}{\partial r})
[(\partial _{-}t)(\partial _{+}\phi )+
(\partial _{+}t)(\partial _{-}\phi )]-
(\frac{\partial R}{\partial r})
(\partial _{+}t)(\partial _{-}t)
\eea 
The constraints become
\bea 
-R(\partial _{\pm }t)^{2}&+&2\alpha (R-1)
(\partial _{\pm }t)(\partial _{\pm }\phi )+\nn \\
&+&[r^{2}+2\alpha ^{2}-\alpha ^{2}R](\partial _{\pm }\phi )^{2}+
\frac{r^{2}}{\alpha ^{2}+r^{2}R}(\partial _{\pm }r)^{2}=0
\eea 
It is straightforward to verify that these constitute the same set 
of equations to be satisfied with the set of 
Eqs (2)-(4) when one sets $\; \th =(\pi /2)\; $.
The action functional, the equations of motion 
and the constraints are invariant under the residual gauge symmetry
\be 
\chi ^{\pm }\Rightarrow \tilde {\chi ^{\pm }}=f_{\pm }(\chi ^{\pm }) 
\ee 
where $\; \; f_{\pm }\; \; $are arbitrary functions of the 
respective arguments.

\section*{IV. Classes of solutions}\

{\Large {\bf i.}} $\; \; $We introduce the ansatz
\be 
(\partial _{\pm }\phi )=
\frac{\alpha (1-R)}{(r^{2}+2\alpha ^{2}-\alpha ^{2}R)}
(\partial _{\pm }t) 
\ee 
Equation of motion (27) is satisfied identically. 
Substituting into the constraints we obtain, along with the previous 
relation the set of equations
\bea 
(\partial _{\pm }t)=\eps (r^{2}+2\alpha ^{2}-\alpha ^{2}R)^{1/2}
\frac{r}{(\alpha ^{2}+r^{2}R)}
(\partial _{\pm }r)\\ 
(\partial _{\pm }\phi )=
\frac{\alpha \eps (1-R)}{(r^{2}+2\alpha ^{2}-\alpha ^{2}R)^{1/2}}
\frac{r}{(\alpha ^{2}+r^{2}R)}
(\partial _{\pm }r)
\eea 
with $\; \; \eps =\pm 1 \; \; $ for expanding or collapsing solutions, 
as observed by an asymptotic observer. 
There exists no need for 
integrability conditions for $\; \; t,\phi \; \; $, because, 
due to the form of eqs (32) and (33), they are  
satisfied identically. 
Substituting these relations into the two remaining equations of
motion (26) and (28) we obtain the same equation to be satisfied by 
$\; \; r=r(\chi ^{+},\chi ^{-})\; \; $,
\be 
\partial _{+}[\frac{r}{(r^{2}+2\alpha ^{2}-\alpha ^{2}R)^{1/2}}
(\partial _{-}r)]+
\partial _{-}[\frac{r}{(r^{2}+2\alpha ^{2}-\alpha ^{2}R)^{1/2}}
(\partial _{+}r)]=0 
\ee 
This can be written as a wave equation 
$\; \; \partial _{+}\partial _{-}f(r)=0 \; \; $ with
\be 
\frac{df(r)}{dr}=\frac{r}{(r^{2}+2\alpha ^{2}-\alpha ^{2}R)^{1/2}}>0 
\ee 
The solution of the wave equation together with the periodicity requirement
for closed strings gives
\bea
f(r)&=&f+g\alpha ^{'}[\chi ^{+}+\chi ^{-}]+\nn \\
&+&i\sqrt{\alpha ^{'}}\sum _{n\neq 0}
\frac{1}{n}[f_{n}e^{-in\chi ^{-}}+
\tilde{f}_{n}e^{-in\chi ^{+}}] 
\eea 
with $\; \; f, g, f_{n}, \tilde{f}_{n} \; \; $constants,  
satisfying the reality conditions for the solution , 
$\; \; f_{-n}=f^{*}_{n}\; \; $, and
$\; \; \tilde{f}_{-n}=\tilde{f}^{*}_{n}\; \; $.

From this we obtain 
\be 
\frac{r\dot{r}}{(r^{2}+2\alpha ^{2}-\alpha ^{2}R)^{1/2}}
=2g\alpha ^{'}+\sqrt{\alpha ^{'}}
\sum _{n\neq 0}[f_{n}e^{-in\chi ^{-}}+\tilde{f}_{n}e^{-in\chi ^{+}}] 
\ee 

The two dimensional worldsheet is in fact a null hypersurface. This is 
easily seen, if we substitute into the relation for the induced metric
\be 
h_{\alpha \beta }=G_{AB}(X)\partial _{\alpha }X^{A}
\partial _{\beta }X^{B} 
\ee 
We can state equivalently that the invariant string size [2], vanishes.
\be 
ds^{2}=G_{AB}(X)\partial _{+}X^{A}
\partial _{-}X^{B}(d\tau ^{2}-d\sigma ^{2})=0 
\ee 

The quantity in the square root is non-negative. Indeed 
for $\; \; \; (R_{-}\leq r \leq R_{+})\; \; $, we have
$\; \; R(r)\leq 0 \; \; $ so 
this is evident while
for $\; \; \; (R_{+}\leq r)\; \; $ we have for 
$\; \; \; M^{2}\geq (\alpha ^{2}+Q^{2})\; \; \; $
\bea 
(r^{2}+2\alpha ^{2}-\alpha ^{2}R)&=&
r^{2}+2\alpha ^{2}-\frac{\alpha ^{2}}{r^{2}}
[r^{2}-2Mr+Q^{2}]=\nn \\
&=&\frac{1}{r^{2}}[(r^{2}+\alpha ^{2})^{2}-\alpha ^{2}
(r^{2}-2Mr+\alpha ^{2}+Q^{2})]\geq \nn \\
&\geq &\frac{1}{r^{2}}[(r^{2}+\alpha ^{2})^{2}-
\alpha ^{2}(r-M)^{2}]\geq 0 \Leftrightarrow \nn \\
&\Leftrightarrow & 
(r^{2}+\alpha ^{2})\geq \alpha \mid r-M\mid 
\Leftrightarrow \nn \\
&\Leftrightarrow & 
(r^{2}+\alpha ^{2})\geq \alpha (r-M) 
\; \; \; (r\geq M) 
\eea 
which holds for $\; \; \; (r\geq R_{+})\; \; \; $. So for 
$\; R_{-}\leq r<+\infty \; $ from eq (25) we have that the 
function f(r) is {\it strictly increasing} in this interval and therefore
{\it invertible}. Denoting the inverse by $\; (\phi )\; $ we have 
$\; r=r(\chi ^{\pm })=\phi [\Phi _{+}(\chi ^{+})+\Phi _{-}(\chi ^{-})]\; $, 
with $\; (\Phi _{\pm })\; $ arbitrary functions of the indicated arguments.
This family of solutions depends on these two arbitrary functions and the two 
arbitrary constants coming from the integration of eqs (22), (23). 

{\Large {\bf ii.}}$\; \; $
We perform the ansatz 
\be 
(\partial _{\pm }t)=\frac{\alpha (1-R)}{(-R)}
(\partial _{\pm }\phi )
\ee 
Equation (26) is satisfied identically. Substituting into the 
constraints, we have non-trivial solutions, provided that 
\bea 
R(r)\leq 0 \Leftrightarrow R_{-}\leq r\leq R_{+}\nn \\
R_{\pm }=M\pm \sqrt{M^{2}-Q^{2}}
\eea 
being the Outer and Inner static limits.
Also we have
\bea 
\rho (r)\equiv \alpha ^{2}+r^{2}R(r)=
\alpha ^{2}+r^{2}-2Mr+Q^{2}\leq 0 
\Leftrightarrow \nn \\
\Leftrightarrow \rho _{-}\leq r\leq \rho _{+}
\; \; \; 
\rho _{\pm }=M\pm \sqrt{M^{2}-Q^{2}-\alpha ^{2}}
\eea 
which are the Inner and Outer Horizons.

From these we obtain the useful approximation relations
\bea 
\rho (r)\simeq \left\{
\begin{array}{ll}
          0 & \mbox{$r\simeq \rho _{+}$}\\
\alpha ^{2} & \mbox{$r\simeq R_{+}$}\\
      r^{2} & \mbox{$r\gg R_{+}$}
\end{array}
\right. 
\eea 
From the first we have that $\; \; r^{2}\simeq (\alpha ^{2}/(-R))\; \;$ so we 
obtain
\bea 
[r^{2}+2\alpha ^{2}-\alpha ^{2}R]\simeq \left\{
\begin{array}{ll}
\frac{(r^{2}+\alpha ^{2})^{2}}{r^{2}} & \mbox{$r\simeq \rho _{+}$}\\
                 (r^{2}+2\alpha ^{2}) & \mbox{$r\simeq R_{+}$}\\
                  (r^{2}+\alpha ^{2}) & \mbox{$r\gg R_{+}$}
\end{array}
\right. 
\eea 

We have therefore the solutions

A)$\; \;  (\rho _{+}\leq r\leq R_{+})$. {\bf Ergosphere}

\bea 
(\partial _{\pm }\phi )=\epsilon 
\frac{r\sqrt{-R}}{\rho (r)}
(\partial _{\pm }r)\\ 
(\partial _{\pm }t)=\epsilon 
\frac{\alpha (1-R)}{\sqrt{-R}}
\frac{r}{\rho (r)}
(\partial _{\pm }r)
\eea 
with $\; \; \epsilon =\pm 1\; \; $ for expanding or collapsing
solutions.

B)$\; \;  (\rho _{-}\leq r\leq \rho _{+})$. {\bf In the Horizon.}
\bea 
(\partial _{\pm }\phi )=\epsilon 
\frac{r\sqrt{-R}}{(-\rho (r))}
(\partial _{\pm }r)\\ 
(\partial _{\pm }t)=\epsilon 
\frac{\alpha (1-R)}{\sqrt{-R}}
\frac{r}{(-\rho (r))}
(\partial _{\pm }r)
\eea 
with $\; \; \epsilon =\pm 1\; \; $ for expanding or collapsing
solutions.

In both cases substitution into the equations (27) and (28)
results after straightforward calculations, to the same 
equation to be satisfied by 
$\; \; r=r(\chi ^{+},\chi ^{-})\; \; $, 
which is,  
\be 
\partial _{+}[\frac{r}{\sqrt{-R}}(\partial _{-}r)]
+\partial _{-}[\frac{r}{\sqrt{-R}}(\partial _{+}r)]=0 
\ee 
This can be written as a wave equation
\be 
\partial _{+}\partial _{-}f(r)=0\; \; \; R_{-}\leq r\leq R_{+} 
\ee 
where 
\bea 
f(r)&=&-[2R_{-}+\frac{1}{2}(r-R_{-})+\frac{3}{4}(R_{+}-R_{-})]
\sqrt{(R_{+}-r)(r-R_{-})}+\nn \\
&+&[\frac{3}{4}(R_{+}-R_{-})^{2}+2R_{-}(R_{+}-R_{-})+2(R_{-})^{2}]
Arctan[\sqrt{\frac{(r-R_{-})}{(R_{+}-r)}}]
\eea 
Again the function f(r) is invertible and the class depends on two 
arbitrary integration functions and two constants.
This class of solutions also gives a null hypersurface as in the 
first case from eqs (38) and (39).

{\Large {\bf iii.}} $\; \; $We introduce the ansatz
\bea 
(\partial _{-}\phi )=\frac{1}{(\alpha ^{2}+r^{2}R)}
(\partial _{-}\Phi )\\ 
(\partial _{+}\phi )=\frac{-1}{(\alpha ^{2}+r^{2}R)}
(\partial _{+}\Phi )\\ 
(\partial _{-}t)=\frac{(r^{2}+\alpha ^{2})}
{\alpha (\alpha ^{2}+r^{2}R)}
(\partial _{-}\Phi )\\ 
(\partial _{+}t)=\frac{-(r^{2}+\alpha ^{2})}
{\alpha (\alpha ^{2}+r^{2}R)}
(\partial _{+}\Phi )\\ 
\alpha (\partial _{-}r)=
(\partial _{-}\Phi )\\ 
\alpha (\partial _{+}r)=-
(\partial _{+}\Phi )
\eea 
The equations of motion (26)-(28) and the constraints (29) are satisfied 
identically, provided that $\; \; \; \Phi =\Phi (\chi ^{+},\chi ^{-})\; \; $ 
is a solution of the wave equation 
$\; \; \; \partial _{+}\partial _{-}\Phi =0\; \; \; $. \\
Again the integrability conditions for the three functions are 
satisfied by the combined use of the wave equation for 
$\; \; \; \Phi =\Phi (\chi ^{+},\chi ^{-})\; \; \; $ and eqs 
(57)-(58). All the three classes of solutions represent 
tensionfull strings. However, they also satisfy the null-tensionless 
string eqs (7)-(9) provided that we insert the small dimensionless 
parameter $\; c^{2}=2\lm T\; $ of the perturbation expansion into 
the terms that contain derivatives with respect to the 
$(\sigma )$ and let $\;c\rightarrow 0\; $ [4]. For example 
we will have $\; (\partial ^{2}_{\tau }-c^{2}\partial ^{2}_{\sigma })
\Phi (\chi ^{\pm })=0\; $. Thus, although they have null worldsheet 
manifold, they satisfy the null string equations of motion and 
constraints, only when their small tension limit is taken explicitly. \\
\section*{V. The Extraction of Energy}\
In the Boyer-Lindquist coordinate system 
$(t,r,\theta ,\phi )$, the Killing vector fields are given by
$\; \; \xi ^{\mu }_{(t)}=\delta ^{\mu }_{t} \; \; $ and 
$\; \; \xi ^{\mu }_{(\phi )}=\delta ^{\mu }_{\phi } \; \; $.

The energy-density along the string is given by [5],[13]
\be 
U\equiv -{\bf p}\cdot \xi _{(t)}=-G_{\alpha \beta }
p^{\alpha }\xi ^{\beta }_{(t)}=-p_{t}
\ee 
where $\; \; p_{\alpha }=(\partial {\cal L}/\partial \dot{X}^{\alpha })\; \; $.

We assume that at $\; \; (\tau =0)\; \; $, the string is at spatial
infinity and as the class of solution implies, 
every point of it moves along a null curve. So 
since the radius of the string decreases, we must 
have 
\bea 
r[\tau =0,\sigma ]=+\infty \\ 
\dot{r}[\tau =0,\sigma ]\equiv p_{0}(\sigma )\leq 0\\ 
U[\tau =0,\sigma ]=\mid p_{0}(\sigma )\mid 
\eea 
Equation (62) is the special-relativistic formula 
$\; \; E^{2}-p^{2}=0\; \;$ for massless particles, or photons.
Computing this, using the first class of solutions we get for the
energy of the string, $\; \; (\dot{r}(\tau ,\sigma )\leq 0)\; \; $, 
\be 
U(\tau ,\sigma )=[
\frac{-2r\dot{r}}
{(r^{2}+2\alpha ^{2}-\alpha ^{2}R)^{1/2}}] 
\ee 
However the canonical momenta used above have been defined with
the ambiguity of the above Lagrangian density . This is because 
(see [12] p.553) 
if $\; \; {\cal L},\tilde{{\cal L}}\; \; $ are two Lagrangian densities 
connected by 
\be 
\tilde{{\cal L}}={\cal L}+\frac{\partial }{\partial \tau }
\lm (\phi ,\phi _{,\al },\tau ,\sigma ) 
\ee 
where we denote all the ``fields" collectively by $\; (\phi )\; $, 
then the dynamics that steming from eq (24) is unaltered. 
This is valid {\sl provided} that in the variation, 
in addition to $\; (\delta \phi )_{boundary}=0\; $
one requires that the solutions of our initial Lagrangian $\; {\cal L}\; $
also satisfy $\; (\delta \phi _{,\al })_{boundary}=0\; $. 
But from eq (37) this occurs for this subclass 
when we set $\; f_{n}=\tilde{f}_{n}=0\; $because 
then the variation of $\; (\dot{r})\; $ and consequently from eqs (32)-(33)
of all the other field derivatives are proportional to the variation of (r) 
and therefore vanish at the boundary, 
$\; (\delta \phi _{,\al })_{boundary}=0\; $. 
We shall retain the constants $\; f_{n},\tilde{f}_{n}\; $ for the sake of 
generality only and set them equal to zero at the end. 
This {\it gauge} freedom allows one to 
write the energy properly, 
by choosing the function $\; (\lm )\; $,as 
\bea 
E(\tau )&\equiv &\int _{0}^{2\pi }d\sigma \tilde{U}(\tau ,\sigma )=\nn \\
&=&\int _{0}^{2\pi }d\sigma \left [
2r\dot{r}[\frac{1}{(r^{2}+2\alpha ^{2})^{1/2}}-
\frac{1}{(r^{2}+2\alpha ^{2}-\alpha ^{2}R)^{1/2}}]-p_{0}(\sigma )\right ]
\eea 
Using eqs (59), (63), (64) and (65), the function $\; (\lm )\; $ 
has to satisfy 
\be 
\frac{\partial ^{2}\lm }{\partial \tau \partial \dot{t}}
=p_{0}(\sigma )-\frac{2r\dot{r}}{(r^{2}+2\al ^{2})^{1/2}} 
\ee 
which is directly integrable for $\; (\lm )\; $ giving  
$\; \lm =p_{0}(\sigma )\tau \dot{t}-2(r^{2}+2\al ^{2})^{1/2}\dot{t}\; $. 

The criterion that enforces this choice is twofold. 
First, the first term in eq (65) is introduced 
to correctly reproduce the asymptotic
form of the energy. Indeed as $\; \; r=r(\tau ,\sigma )
\rightarrow +\infty \; \; $ the energy of the string 
must irrespectively of the functional form of 
$\; \dot{r}(\tau ,\sigma )\leq 0\; $,
assume the 
required value (see eq (72) below ). This is satisfied by this 
choice since we now have 
$\; \tilde{U}(\tau =0,\sigma )=\mid p_{0}(\sigma )\mid \; $.

Secondly 
one sees that orbits of negative energy exist in a region somewhat larger 
than the ergosphere $\; \; r(\tau ,\sigma )\leq R_{+}\; \; $, because the 
factor $\; R(r)\; $in the second term becomes negative. 
This is in 
conformity with what one expects in analogy with the particle case. 

So we examine the case that the string decays into two parts and 
one of them asymptotically $\; \; (t\rightarrow +\infty )\; \; $,
as observed by the asymptotic observer, 
reaches a stable, negative-at infinity energy state [13]. 
Indeed this hapens because, for $\; r\simeq \rho _{+}\; $, from eq (32) 
we have that 
\be 
\left.\frac{dr}{dt}\right|_{r\simeq \rho _{+}}\simeq 
\frac{\al ^{2}+\rho _{+}^{2}R(\rho _{+})}
{\rho _{+}(\rho _{+}^{2}+\al ^{2})^{1/2}}=0 
\ee  
using eq (43).

From eq (37) we have, using the above notation conventions, 
that substitution of eq (61) gives 
\be 
p_{0}(\sigma )=2g\alpha ^{'}+\sqrt{\alpha ^{'}}
\sum _{n\neq 0}[f_{n}e^{+in\sigma }+\tilde{f}_{n}
e^{-in\sigma }] 
\ee 
and from these
\bea 
g=\frac{1}{4\pi \alpha ^{'}}\int _{0}^{2\pi }d\sigma p_{0}(\sigma )\\ 
f_{n}=\frac{1}{2\pi \sqrt{\alpha ^{'}}}
\int _{0}^{2\pi }d\sigma p_{0}(\sigma )e^{-in\sigma }\\ 
\tilde{f}_{n}=\frac{1}{2\pi \sqrt{\alpha ^{'}}}
\int _{0}^{2\pi }d\sigma p_{0}(\sigma )e^{+in\sigma }
\eea 
The energy of the string at infinity is given
\be 
E_{i}=E(\tau =0)=-\int _{0}^{2\pi }d\sigma p_{0}(\sigma ) 
\ee 
and the energy at later times when it approaches a stable 
negative-at infinity  
energy state
by
\bea
E(\tau )=\int _{0}^{2\pi }d\sigma
\left[
2\left[\frac{(r^{2}+2\alpha ^{2}-\alpha ^{2}R)^{1/2}}
{(r^{2}+2\alpha ^{2})^{1/2}}-1\right]\right.\cdot \nn \\
\cdot \left.[2g\alpha ^{'}+\sqrt{\alpha ^{'}}\sum _{n\neq 0}
(f_{n}e^{-in\chi ^{-}}+\tilde{f}_{n}e^{-in\chi ^{+}})]
-p_{0}(\sigma )\right]
\eea 
where use of eq (37) has been made.\ 
We now introduce the previous result that for this 
orbit we have with high approximation 
$\; \; \; r\simeq \rho _{+}\; \; \; $ and set 
$\; f_{n}=\tilde{f}_{n}=0\; $. 
Factoring out the term with the (r) dependance and 
performing the integral, we obtain
for the energy in a stable negative-energy state, 
with the help of the relation 
$\; r^{2}\simeq (\al ^{2}/(-R))\; $ 
\be 
\frac{E(stable)}{E(\tau =0)}=1-2\left[
\frac{(\rho _{+}^{2}+\alpha ^{2})}{\rho _{+}
\sqrt{\rho _{+}^{2}+2\alpha ^{2}}}-1
\right]<1  
\ee 
This means that the energy of the string as measured asymptotically has been 
reduced from its initial value due to the trapping in the negative energy 
state. 
Now if the string decays at two parts with a proportionality 
factor $\; \; \bar{\lambda }
 \; \; (0<\bar{\lambda }<1)\; \; $ where the $\; (\bar{\lambda })\; $ 
part is trapped in the negative energy state, while the 
$\; (1-\bar{\lm })\; $ part escapes at infinity, 
then the energy that one obtains with respect to the energy that 
was thrown in the black-hole, is given by the efficiency ratio, 
\be 
{\cal \epsilon }\equiv \frac{E_{f}}{E_{i}}=1+2\bar{\lm }
\left[
\frac{(\rho _{+}^{2}+\alpha ^{2})}{\rho _{+}
\sqrt{\rho _{+}^{2}+2\alpha ^{2}}}-1\right]>1 
\ee 
This occurs as follows. The second term on the r.h.s. of eq (74) 
is negative and so, for the $\; (\bar{\lm })\; $ part of the string 
it represents the excess energy that is obtained asymptotically. 
One can verify in eq (75) that the quantity in the brackets is positive. 
Expanding the square root one sees that the remaining term is of the order 
of $\; (\al ^{4})\; $. 
When no decay occurs ($\; \;\bar{\lm }=0 \; \; $) we have, 
as expected, equality of the emited and received, energies.
Also when the black-hole is non-rotating, $\; \; \; (\alpha =0)\; \; $, 
we have the same result, that is one obtains no energy gain. 

\section*{VII. Discussion}\ 
Three classes of solutions for null strings, that reside on the equatorial
plane of a Kerr-Newmann black hole were presented. All of them depend on two
arbitrary integration functions and two integration constants, 
coming from wavelike differential 
equations.

From the first of these classes, an efficiency coefficient for the
energy extraction was calculated, for a closed string that 
encircles the black hole and decays into
two parts, one of which is trapped at a stable, negative-at infinity,  
energy state [13-14].

It would be interesting to generalize this argument to the case where the 
string is not bound to move on the equatorial plane. Also an important case 
is to consider bosonic strings in their classical as well their quantum 
description. Of the same importance is to examine the case where one has initially 
a stationary axisymmetric metric, try to obtain cosmic string solutions 
of solitonic nature and examine their properties [15]. 
This however is extremely difficult due to the abscence of 
exact solutions in the general case. Also it would be very interesting to 
consider more general extended objects such as null bosonic p-branes 
and examine similar energy-extraction processes for these spacetime 
backgrounds [16]. 
Work along these directions is in progress 
and will reported promptly. 

\section*{Acknowledgements} \
The authors would like to thank Professor Louis Witten for valuable
discussions. \\
The authors would like to thank Professor N. K. Spyrou for valuable 
discussions. \\
The authors would like to thank Professor Valeri Frolov for providing 
the exact form of the Refs. [5-7]\\
One of us (A.K) would like to thank the Greek
State Scholarships Foundation (I.K.Y), for the financial support 
during this work.\\ 
The authors would like to thank Dr. K. Kleidis for valuable discussions.\\ 
Financial support from the Greek Secretariat of Research and Technology, 
(PENED) under contract (1768) is gratefully acknowledged.\\

\end{document}